\documentclass[conference]{IEEEtran}

\ifCLASSINFOpdf
   \usepackage[pdftex]{graphicx}
\else
   \usepackage[dvips]{graphicx}
\fi
\usepackage[cmex10]{amsmath}
\usepackage{rotating}

\usepackage{array}                                                                       
\usepackage{amsmath}

\usepackage{booktabs}
\usepackage{ulem}
\usepackage{float}
\floatstyle{ruled}
\newfloat{model}{thp}{lop}
\floatname{model}{\textsc{Model}}
\usepackage{bm}
\usepackage{cite}
\usepackage[colorlinks=true,citecolor=blue,linkcolor=blue,urlcolor=blue,plainpages=false]{hyperref}
\usepackage{nomencl}
\usepackage{cleveref}

\makenomenclature

\usepackage{etoolbox}
\renewcommand\nomgroup[1]{%
  \item[\bfseries
  \ifstrequal{#1}{I}{\textit{Indices and Sets}}{%
  \ifstrequal{#1}{P}{\textit{Parameters}}{%
  \ifstrequal{#1}{F}{\textit{Functions}}{%
  \ifstrequal{#1}{M}{\textit{Production Models}}{%
  \ifstrequal{#1}{V}{\textit{Variables}
  }{}}}}}%
]}

\newcommand{\Pcc}{\text{Pcc}}
\def\R{{\bm{R}}}

\IEEEoverridecommandlockouts

\makeatletter
\renewcommand*{\fnum@model}{\fname@model}

\makeatother

\begin{document}

\title{Construction of Multi-period TSO-DSO Flexibility Regions\\
\thanks{This project was funded by the Skolkovo Institute of Science and Technology as a part of the Skoltech NGP Program (Skoltech-MIT joint project).}
}

\author{
\IEEEauthorblockN{Luis Lopez, Alvaro Gonzalez-Castellanos, David Pozo}
\IEEEauthorblockA{Center for Energy Science and Technology \\
Skolkovo Institute of Science and Technology (\textit{Skoltech})\\
Moscow, Russian Federation\\
}
}

\maketitle

\begin{abstract}
Active distribution networks (ADN) have grown considerably in recent years. Distributed energy resources present in ADNs can provide flexibility to the power system through TSO/DSO coordination, i.e.,  at the interface node (feeder) between the transmission and distribution network. This paper addresses the issue of calculating multi-period flexibility regions of the ADNs. Flexibility regions are tightly dependent between periods and conditioned on the actual deployment of such flexibilities in real-time. 
The existing state-of-the-art has not provided a robust methodology for building multi-period flexible regions. 
We present a new mathematical framework based on a non-iterative formulation that considers the multi-period flexibility boundary points in a single optimization problem. The proposed methodology is evaluated on IEEE standard test networks and compared with the most widely used methods in the literature.
\end{abstract}

\begin{IEEEkeywords}
Distributed energy resources, Flexibility, Multiperiod constraints, TSO/DSO interface, Active distribution grid. 
\end{IEEEkeywords}


\printnomenclature[0.65in]

\section{Introduction}

Energy policies for reducing CO2 emissions are promoting the penetration of renewable energy sources. But the uncertainty of renewable sources and the decommitment of conventional plants creates a lack of flexibility in power systems \cite{rogge2017conceptual}. One possible solution envisioned by the scientific community and industry sector for facing this problem is to exploit the flexibility that distributed energy resources (DER) can provide. However, integrating a large number of flexibility elements into electricity markets is an important matter, albeit a challenging  task \cite{Bolfek2021}. Some progress has been made. For instance, countries like Sweden and Finland support legislation where the DSO manages flexibility, but the TSO generally manages system security \cite{Silva2018}. Thus, TSO-DSO coordination arises as an appealing solution for the aggregation of DERs to provide network flexibility services from distribution  to transmission grids \cite{ageeva2020coordination}.

TSO-DSO coordination is a challenging task due to the large number of flexible resources connected. In recent years, grid-connected DERs have expanded rapidly, such as electric vehicles and inverter-based distributed generators (e.g., PV power plants and micro wind turbines). Electricity markets (such as Australia) have studied the need and urgency of DER integration to take advantage of their flexibility potential \cite{Givisiez2020a}. Flexibility estimation provided at the DSO interface should be studied at both the technical and regulatory levels. In this work, we focus on the technical level. The flexibility estimation at the boundary node of the two operators is one input of the TSO-DSO coordination. Identifying the flexibility ranges, the DSOs can support transmission network decisions and provide ancillary services \cite{burger2017business, nosratabadi2017comprehensive}.

\subsection{Literature review}

The potential in the interaction of distribution and transmission operators has generated substantial research in this field. To date, different computational methods have been proposed to estimate the feasible operating region at the TSO-DSO interface \cite{contreras2021}. Methods based on random sampling (Monte Carlo), point sum of feasible regions (Minkowsky) and algorithms based on iterative optimization are the methodological directions employed to aggregate operational flexibility in distribution networks.

The Minkowski technique is a good approximation of aggregated flexibility in the absence of network constraints. However, it does not consider operational limitations, such as line and voltage limits, so there is no accurate representation of the available flexibility. This technique can lead to an overestimate of the flexibility region \cite{silva2018challenges}. 

In the Monte Carlo technique, power generation instances are run randomly; then, the infeasible points are discarded, and the remaining points represent the flexibility region. Although the Monte Carlo method is simple, it is limited by the large number of points required to obtain an accurate estimate \cite{contreras2021}. 

The optimization techniques approach aims to find the flexibility region by considering network and DERs constraints. However, the size and complexity of the optimization problem grow with the size of the network and the number of DERs and can be computationally demanding \cite{lopez2021quickflex}. 

\subsection{Paper contributions and organization}

The core contribution of this paper lies in the modeling aspects for the estimation of multi-period flexibility regions for TSO-DSO coordination.  
%
In addition, this paper contributes to the growing debate on the coordination between TSO and DSO driven by massive energy resources located in distribution networks. 
%
The main contributions of our work are specifically: 
    We propose a non-iterative formulation for the computation of the multi-period TSO-DSO feasible. We estimate the polygon enclosing the flexibilities in an aggregate form. Contrary to the existing literature, we can provide flexible regions that both robust and not dependant on a reference dispatch. Thus, we guarantee that any realized flexibility deployment path within the generated regions will always be feasible realizable at every period. This fundamental concern is not secured when multi-period flexible regions are generated using a reference power dispatch \cite{Riaz2019}. 

The paper is organized as follows: \Cref{Sec:Model} describes the modeling framework used, explains the definition of flexibility, and presents the model formulation. \Cref{Sec:Numerical-analysis} shows the results obtained from evaluating the proposed model on test distribution networks. \Cref{Sec:Discussion} discusses multi-period flexible regions and their impact on TSO-DSO coordination and \Cref{Sec:Conclusions} concludes.

\nomenclature[I]{$t, T$}{Index/set of time periods.}
\nomenclature[I]{$i, N$}{Index/set of nodes.}
\nomenclature[I]{$(i,j), E$}{Index/set of lines.}
\nomenclature[I]{$h, H$}{Index/set of extreme points.}

\nomenclature[P]{$P^D_{i,t}, Q^D_{i,t}$}{Active/reactive power demand in $t$ at bus $i$.}
\nomenclature[P]{$R_{i,j}, X_{i,j}$}{Resistance/inductance of the line from bus $i$ to $j$.}
\nomenclature[P]{$\overline{V}_i,\underline{V}_i$}{Voltage magnitude limits for bus $i$.}
\nomenclature[P]{$\overline{l}_{ij}$}{Current limit in line from bus $i$ to $j$.}
\nomenclature[P]{$\overline{P}^G_i, \underline{P}^G_i$}{Active power limits for generation at bus $i$.}
\nomenclature[P]{$\overline{Q}^G_i, \underline{Q}^G_i$}{Reactive power limits for generation at bus $i$.}
\nomenclature[P]{$\alpha_h$}{Exploration angle direction at $h$.}
\nomenclature[P]{$R^{up}_i,R^{dn}_i$}{Ramp up and ramp down limits.}
\nomenclature[P]{$\Delta t$}{Time interval.}
\nomenclature[P]{$\eta^{C}_i, \eta^{D}_i$}{Charging/discharging efficiency of battery.}
\nomenclature[P]{$\overline{P}^{BC}_i, \overline{P}^{BD}_i$}{Charging/discharging power limits for battery.}
\nomenclature[P]{$\overline{E}_i$}{Capacity limit for energy storage}

\nomenclature[V]{$v_{i,h,t}$}{Voltage magnitude at bus $i$.}
\nomenclature[V]{$l_{ij,h,t}$}{Current magnitude at line from $i$ to $j$.}
\nomenclature[V]{$p_{ij,h,t}$}{Active power flow at line from $i$ to $j$.}
\nomenclature[V]{$q_{ij,h,t}$}{Reactive power flow at line from $i$ to $j$.}
\nomenclature[V]{$p^G_{i,h,t}$}{Active power for generation at bus $i$ in  $t$.}
\nomenclature[V]{$q^G_{i,h,t}$}{Reactive power for generation at bus $i$ in  $t$.}
\nomenclature[V]{$e_{i,h,t}$}{Energy stored in battery at bus i.}
\nomenclature[V]{$p^{BC}_{i,h,t}, p^{BD}_{i,h,t}$}{Battery charging/discharging power input/output.}
\nomenclature[V]{$uc_{i,h,t}$}{Binary variable for charging state of
battery.}
\nomenclature[V]{$ud_{i,h,t}$}{Binary variable for discharging state of
battery.}

\section{Modeling Framework} \label{Sec:Model}

\subsection{Definition of the TSO-DSO flexibility region}

TSOs and DSOs are responsible for the operation of the transmission and distribution networks, respectively. 
DSOs can provide flexibility to TSOs from the aggregated coordination of a set of local resources such as battery storage systems, demand response, electric vehicles, etc. 
That flexibility can be in the form of energy for trading in power markets or providing ancillary services for supporting the transmission grid. 
%

\textit{A TSO-DSO flexibility region is defined as the active (P) and reactive (Q) power capacity area at the TSO-DSO interface, resulting into a realizable distribution grid operation}. TSO-DSO flexibility regions represent additional degrees of freedom for the transmission grid operation. 
From the mathematical point of view, flexibility represents feasibility. In the context of mathematical modeling of TSO-DSO flexible regions, both terms are equivalent. We use both terms along the text to refer to the same object.   



\subsection{Preliminary considerations}

\begin{figure}[htbp]
    \centering
    \includegraphics[width=0.8\columnwidth]{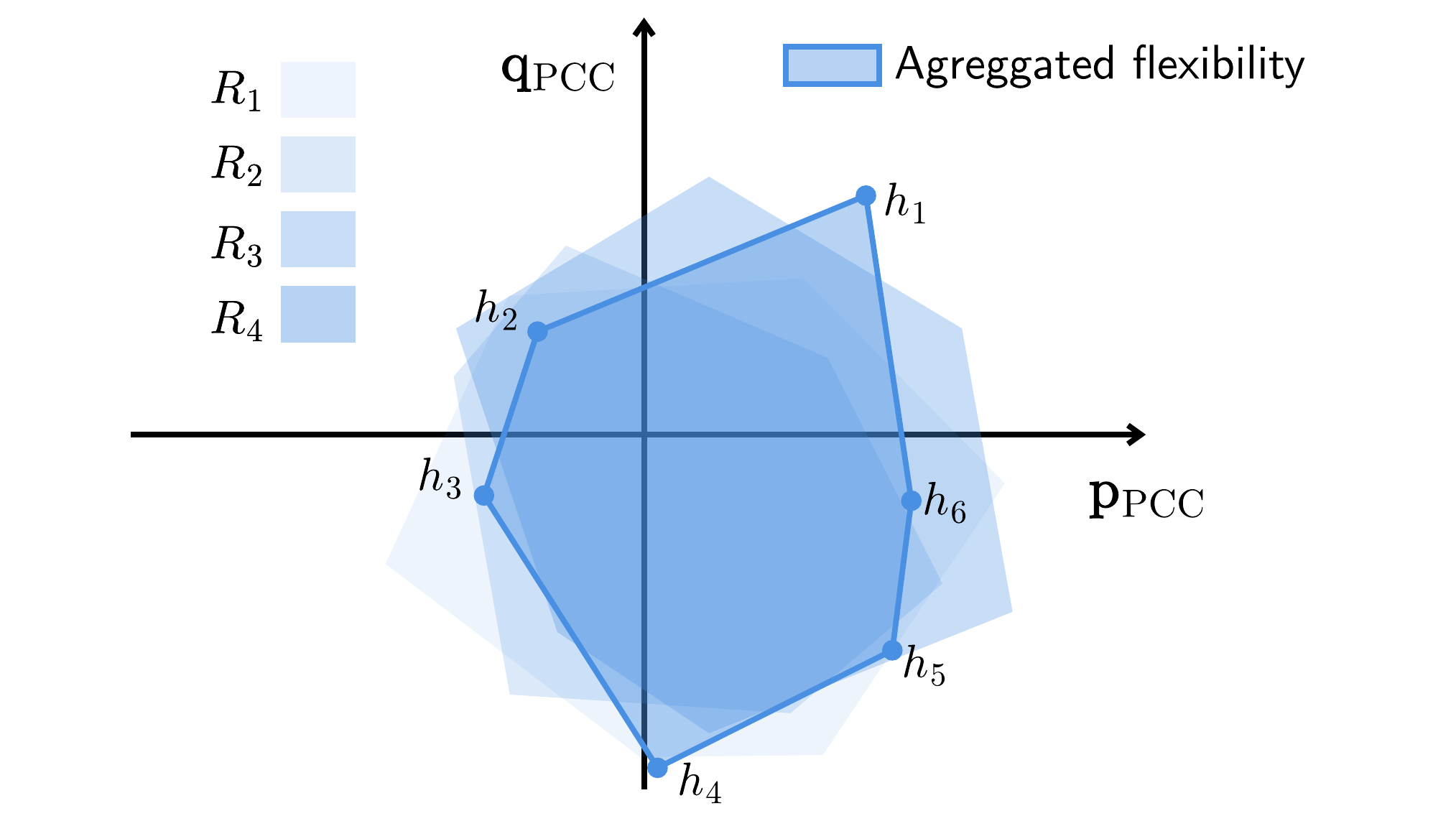}
    \caption{Multi-period flexibility region nomenclature. Example with $6$ extreme points and $4$ periods.}
    \label{fig:Flex_region_nomen}
\end{figure}

The flexible TSO-DSO region at a particular time period $t$ is defined in the  two-dimensional PQ-space by $\R_t$. As customary by the state-of-the-art, \cite{Riaz2019,ageeva2020coordination, lopez2021quickflex}, we characterize the region $\R_t$ as convex hull typically given by the boundary on the convex hull points, i.e.,  $\R_t = \{(p_t^h,q_t^h), \:  \forall h=1,\ldots, |H|\}$. If the  boundary operational points cannot fully be recovered, $\R_t$ would represent an approximation to the original set. In our proposal, we will provide a predefined number of points that should form the convex hull $\R_t$ for each period $t$. Figure \ref{fig:Flex_region_nomen} illustrates the nomenclature used in our formulation. Flexible region are constructed by the boundary points $h$ for every period $t$. Extreme points $h$ are needed to model the inter-temporal constraints between two periods.  

\subsection{Mathematical Formulation}

We describe the formulation used to calculate the multi-period flexibility region. 
The model constraints are divided into four groups. The power flow constraints are associated with the network modeling. The engineering constraints are associated with the technical limits of the network. The restrictions of distributed flexible resources such as generation-like and storage-like devices. Finally, we describe the objective function implemented to calculate the multi-period region.  

\textbf{Network constraints: 
}Distribution networks are typically operated using radial topology to supply the power demand of their users. There is a substation or feeder is the source of active power and ancillary services for passive distribution grids.  In the active distribution grids, local resources can provide active power and some ancillary services. 
The branch power flow equations are described \eqref{eq:act_power_flow}-\eqref{eq:app_power}. This modeling of the distribution network is known as DistFlow and was first proposed in \cite{baran1989optimal}. 

\textbf{Engineering constraints: 
}The voltage at each node in the system must be kept within operating limits dictated by regulation. The constraint \eqref{eq:voltage_lim} represents the limits of voltage magnitude squared. Typically, the range of allowable voltages is $[0.95, 1.05]$. On the other hand, constraint \eqref{eq:current_lim} represents the maximum permissible current squared on a transmission line. The current limits are defined by the maximum thermal capacity of each line. 

\textbf{Generation-like distributed flexible resources: 
}One of the flexible elements considered in the distribution system is modeled as a distributed generator. The constraints \eqref{eq:act_power_lim}-\eqref{eq:re_act_power_lim} represent the maximum active and reactive generation limits for each generator. While constraint \eqref{eq:ramp_limits} represents the ramping limits between two consecutive periods. Note that the ramp-up and ramp-down limits may be different. 

\textbf{Storage-like distributed flexible resources: 
}Flexible storage resources are modeled as batteries in the distribution system. Constraint \eqref{eq:Batt_SOC} shows the model used for the energy balance using the injected power and the efficiency of the charge/discharge states. Constraints \eqref{eq:Batt_lim}-\eqref{eq:dis_batt_lim} show the maximum stored energy limits and the charge/discharge power limits per period. Finally, constraint \eqref{eq:bin_const} presents the relationship of the binary variables to define the battery state of charge/discharge. 

\textbf{Objective function:
} Deciding what objective function would be appropriate for the construction of the multi-period flexible regions is not straightforward. For instance, some periods could be prioritized for flexibility provision due to several economic- and reliability-related issues. Such criteria should be driven by DSOs' preferences.  Without loss of generality, we select two criteria, namely maximum accumulated PQ area and maximum distance for given direction. 

The first objective function, a natural interpretation of the maximal flexibility, is the maximization of the accumulated PQ areas in all periods. We use the \textit{Surveyor formula} \cite{braden1986surveyor} for calculating the area of a polygon. It is depicted in \eqref{eq:obj_surveyor}.
\begin{IEEEeqnarray}{lll}
f_1(\mathbf{p}_{\Pcc},\mathbf{q}_{\Pcc})  =&  \: \sum_{t \in T} \left|\left(\sum_{h=1}^{H-1} p_{\Pcc,ht} q_{\Pcc,h+1,t}\right) \ldots 
\IEEEnonumber \right. \\
& \qquad + \:p_{\Pcc,Ht} q_{\Pcc,1t} \ldots & 
\IEEEyesnumber
\label{eq:obj_surveyor}
\\ 
& - \left(\sum_{h=1}^{H-1} p_{\Pcc,h+1,t} q_{\Pcc,ht} \right)-p_{\Pcc,1t} q_{\Pcc,H,t} \Bigr|
\IEEEnonumber
\end{IEEEeqnarray}

The second objective function seeks to \textit{maximize P and Q for a given set of directions}  \eqref{eq:obj_linear}. It is inspired on the radial reconstruction method proposed previously for a single-period TSO-DSO flexible region \cite{pisciella2017optimal}. As such, a given set of directions defined by the angle $\alpha_h$ are predefined a priori. The larger the set of directions, the better the approximation to the maximal area objective.  
\begin{IEEEeqnarray}{lll}
f_2(\mathbf{p}_{\Pcc},\mathbf{q}_{\Pcc}) = &  \:\sum_{t \in T} \sum_{h \in H}  \Big( p_{\Pcc,ht} \cos{\alpha_h} + q_{\Pcc,ht} \sin{\alpha_h} \Big) \;\:
\IEEEyesnumber
\label{eq:obj_linear}
\end{IEEEeqnarray}
Notice the objective function \eqref{eq:obj_linear} is linear, but it does not directly maximize the area of the enclosed polygon. In contrast the objective function \eqref{eq:obj_surveyor} does maximize the area but has non-convex components due to the bilinear products between the active and reactive power variables.

\begin{model}[htbp]
\caption{Multiperiod flexibility map estimation \hfill [NLP]}
\label{Mod: DistFlow}

\mbox{\bf Objective: } 
\begin{IEEEeqnarray}{lll}  \label{eq. model_FR}
& \max_{p^G_{\Pcc},q^G_{\Pcc}} \:
f(\mathbf{p}_{\Pcc},\mathbf{q}_{\Pcc}) &
\IEEEyesnumber
\IEEEyessubnumber
\end{IEEEeqnarray}

\mbox{\bf Subject to: }
\begin{IEEEeqnarray}{llr}
\Bigg\{& p_{ijht} = P^D_{jt} - p^G_{jht} + p^{BC}_{iht} - p^{BD}_{iht} ... & 
\IEEEnonumber
\\ 
& \; \qquad \qquad  ...+ R_{ij}l_{ijht} + \sum _{k:(jk) \in E} p_{jk} &
\IEEEyessubnumber
\label{eq:act_power_flow}\\
& q_{ijht} = Q^D_{jt} - q^G_{jht} + X_{ij}l_{ijht} + \sum _{k:(jk) \in E} q_{jk} &
\IEEEyessubnumber
\label{eq:react_power_flow}\\
& v_{jht} = v_{iht} - 2( R_{ij} p_{ijht} + X_{ij} q_{ijht}) ... & 
\IEEEnonumber
\\ 
& \; \qquad \qquad  ...+\left( R_{ij}^{2} +X_{ij}^{2}\right) l_{ijht} &
\IEEEyessubnumber
\label{eq:volt_flow}\\
& l_{ijht} v_{iht} =p_{ijht}^{2} +q_{ijht}^{2} & 
\IEEEyessubnumber
\label{eq:app_power}\\
& \underline{V}_i \leq v_{iht} \leq \overline{V}_i & 
\IEEEyessubnumber
\label{eq:voltage_lim}\\
& l_{ijht} \leq \overline{l}_{ij} &
\IEEEyessubnumber
\label{eq:current_lim}\\
& \underline{P}^G_i \leq p^G_{iht} \leq \overline{P}^G_i&
\IEEEyessubnumber
\label{eq:act_power_lim}\\
& \underline{Q}^G_i \leq q^G_{iht} \leq \overline{Q}^G_i& 
\IEEEyessubnumber
\label{eq:re_act_power_lim}\\
& -R^{dn}_i \leq p^G_{ih,t+1} - p^G_{ih\prime t} \leq R^{up}_i&
\IEEEyessubnumber
\label{eq:ramp_limits}\\
& e_{iht} = e_{ih,t-1} + (p^{BC}_{iht}\eta^{C}_i - \frac{p^{BD}_{iht}}{\eta^{D}_i})\Delta t & 
\IEEEyessubnumber
\label{eq:Batt_SOC}\\
& 0 \leq e_{iht} \leq \overline{E}_i &
\IEEEyessubnumber
\label{eq:Batt_lim}\\
& 0 \leq p^{BC}_{iht} \leq uc_{iht}\overline{P}^{BC}_i &
\IEEEyessubnumber
\label{eq:cha_batt_lim}\\
& 0 \leq p^{BD}_{iht} \leq ud_{iht}\overline{P}^{BD}_i &
\IEEEyessubnumber
\label{eq:dis_batt_lim}\\
& uc_{iht} + ud_{iht} \leq 1 \quad  \quad \quad \Bigg\} \: \forall t\in T, h \in H
\IEEEyessubnumber
\label{eq:bin_const}
\end{IEEEeqnarray}
\end{model}


We want to highlight that the optimization problem \eqref{eq. model_FR} provides the set of all realizable $(\mathbf{p}_{\Pcc},\mathbf{q}_{\Pcc})$ paths at the TSO-DSO interface.  At the same time, the model assumes a deterministic approach for the associated available flexibility from distributed resources. We leave this subject for future work. 

\section{Numerical analysis} \label{Sec:Numerical-analysis}

This section presents the description of the test system, the case studies, and the results obtained.

\subsection{Illustrative example}
\begin{figure}[H]
    \centering
    \includegraphics[width=0.8\columnwidth]{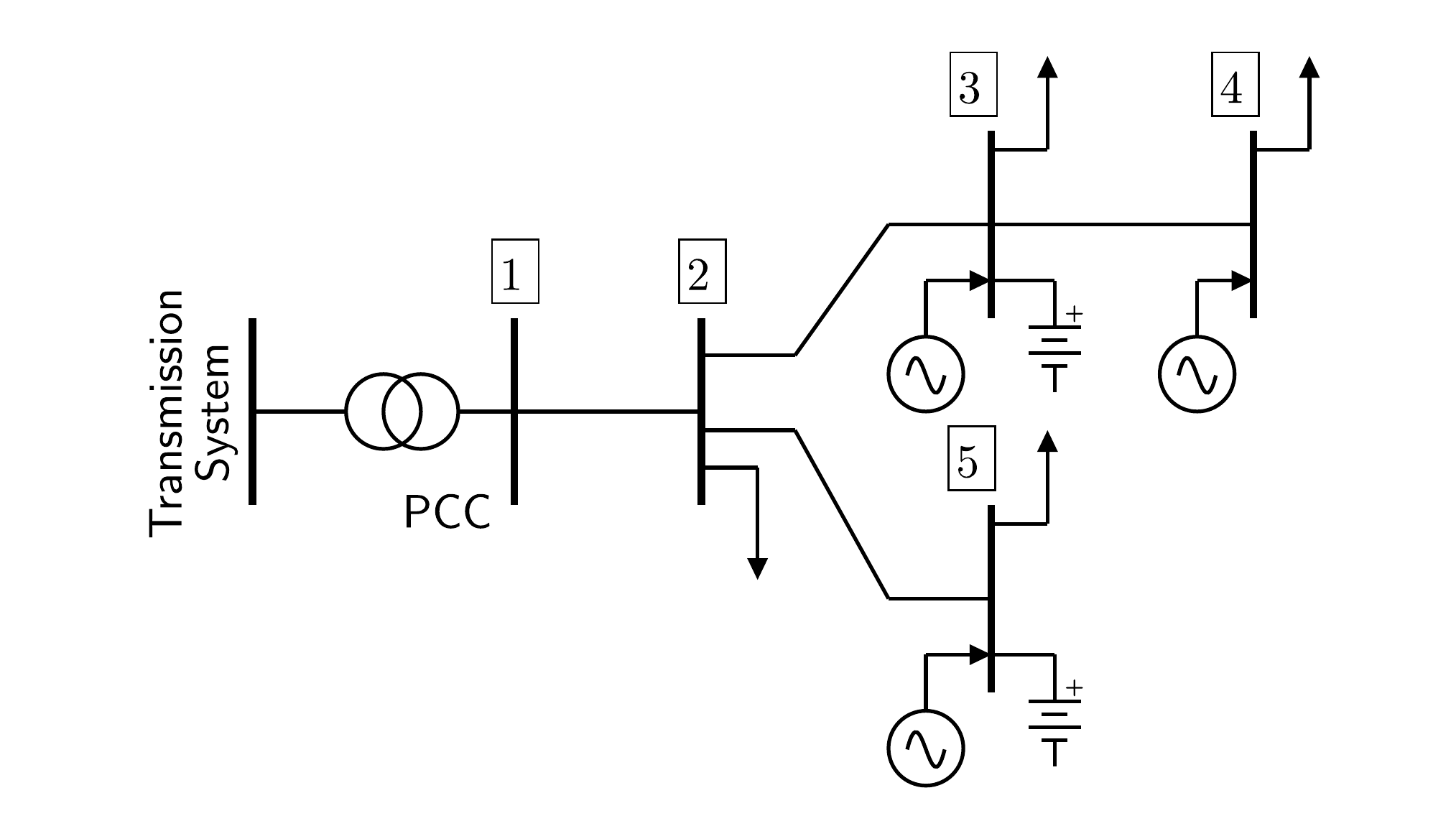}
    \caption{IEEE 5-bus system. }
    \label{fig:Test_example}
\end{figure}

The presented model is evaluated on the modified 5-node system, as shown in \Cref{fig:Test_example}. The network characteristics, such as line impedance and load requirements, are taken from the original model presented in \cite{li2010small}. The network is modified by adding three distributed generators with  a total installed capacity of $1.5$ MW. In addition, two batteries with a total energy capacity of $200$ kWh are added. The net active and reactive power demand is $1.3$ MW and $0.427$ MVAR, respectively. 

\begin{figure*}[bht]
\centering
    \begin{minipage}{0.32\textwidth}
		\centering
    	\includegraphics[width=\textwidth]{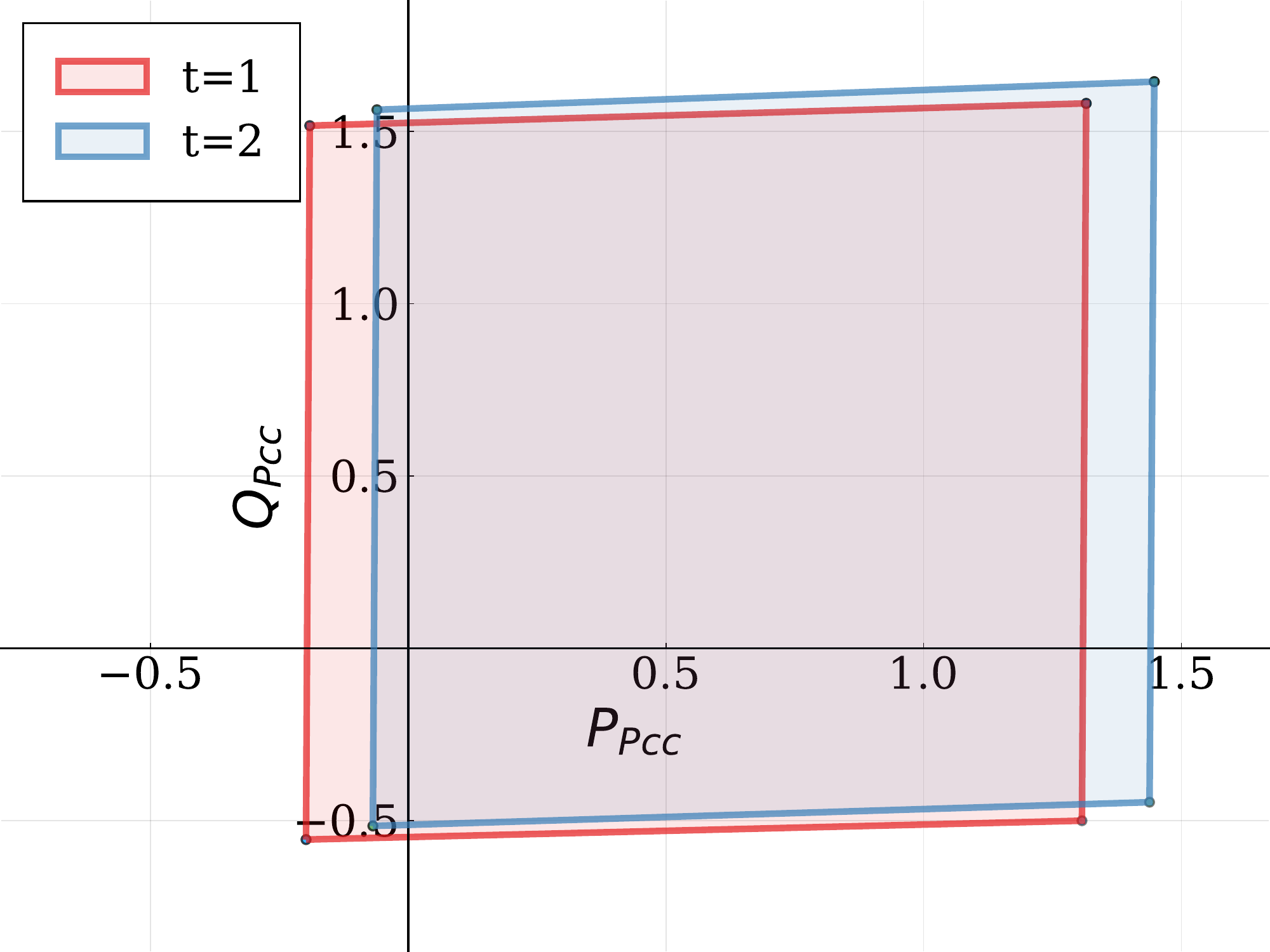}
	\end{minipage}
	\hspace{0.05cm}
	\begin{minipage}{0.32\textwidth}
	\includegraphics[width=\textwidth]{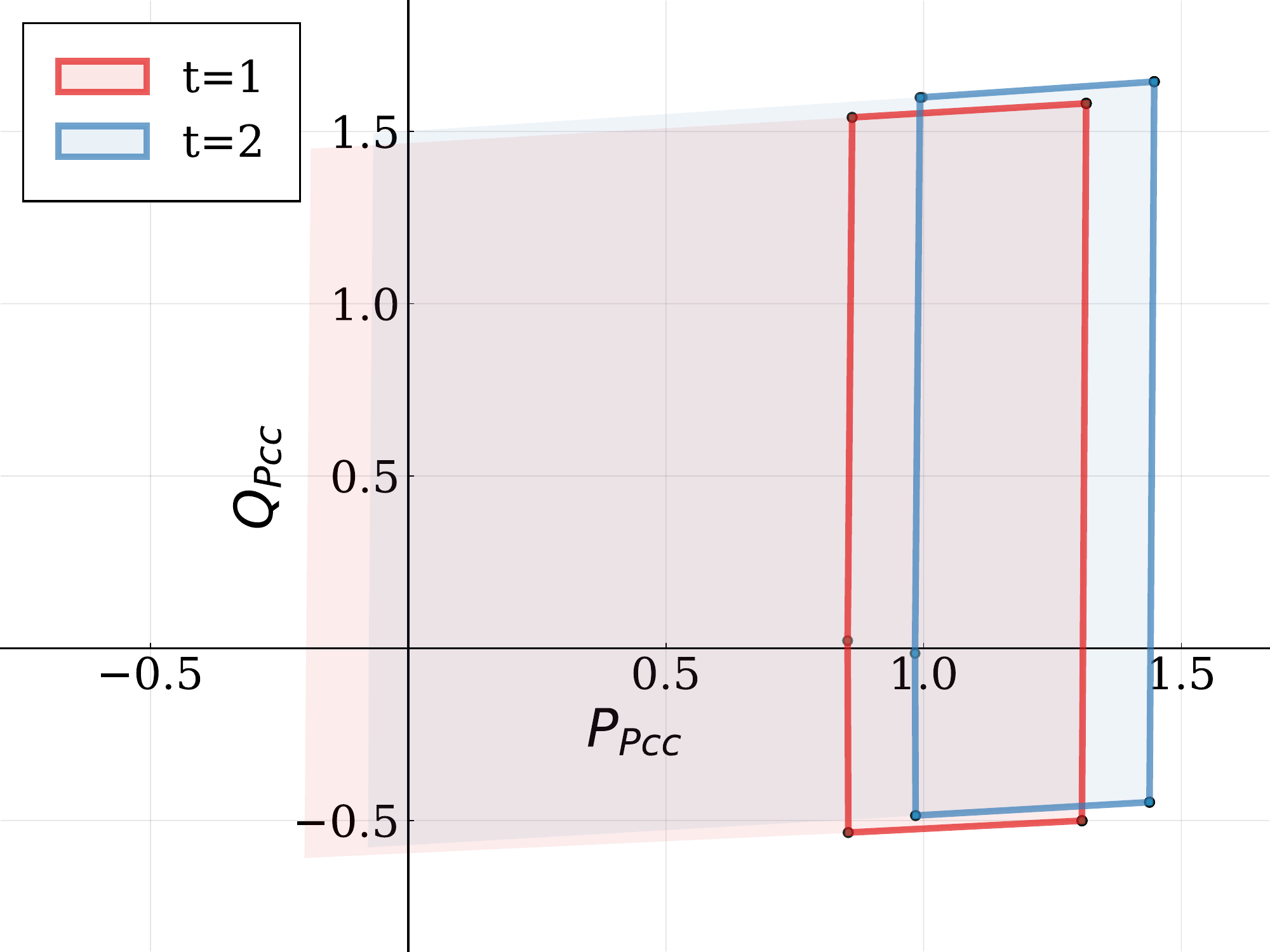}
	\end{minipage}
	\hspace{0.05cm}
	\begin{minipage}{0.32\textwidth}
		 \includegraphics[width=\textwidth]{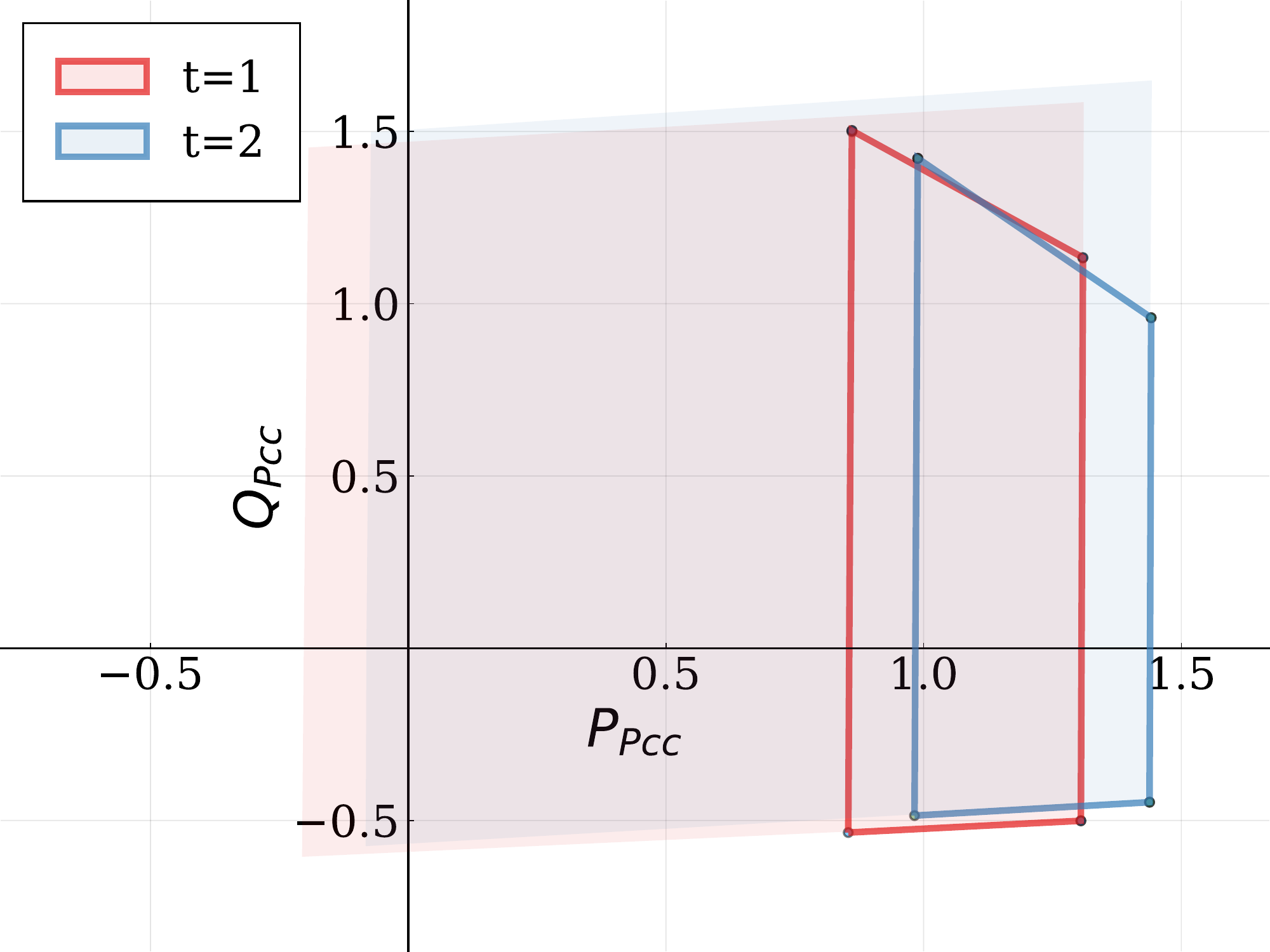}
	\end{minipage}
	\caption{Feasible regions at the TSO/DSO interface with different ramps and distribution networks constrints: (left) no constraints, (center) ramp constraints, and (right) ramp and network constrains.} 	\vspace{-0.5cm}
	\label{fig:five-node-example}
\end{figure*}

\subsection{Test cases}
Four different test cases have been designed to analyze the estimation of the flexible region. 
\begin{itemize}
    \item \textbf{Case I} estimates the flexibility of the distribution system without considering inter-temporal relationships and network constraints. 
    \item \textbf{Case II} considers a ramp for distributed generators of 5\%, 30\%, and 50\% of the total installed capacity. 
    \item \textbf{Case III} considers ramp constraints and network constraints such as thermal limits and voltage limits. 
    \item \textbf{Case IV} considers ramp constraints, storage devices and network constraints such as thermal limits and voltage limits. 
\end{itemize}

For a better understanding of the multi-period flexible region, we have plotted only two periods. The period $t=1$ with the nominal system demand. The period $t=2$ with the system demand multiplied by a factor of $1.1$. The computational experiments were implemented in Julia with JuMP and executed on an Intel(R) Xeon(R) Gold 6148, CPU 2.40GHz, 2394 Mhz, 20 Core(s), 256 GB of RAM.

\subsection{Analysis of flexibility regions}

\Cref{fig:five-node-example} shows the results obtained from the multiperiod region for the 5-node system studied. To plot these flexible regions, we have used Surveyor formula in the objective function. In addition, we have defined a total of 8 points to describe the polygon enclosing the area in the PQ plane. For Case I, where there are no network or intertemporal constraints, the solution corresponds to the possible active and reactive power operating points that the DSO can have. The active power range is related to the maximum demand for that period and to the full power injection that the DERs can deliver. The same is valid for the reactive power limits. 

In the central graph of \Cref{fig:five-node-example} (Case II) a ramp constraint corresponding to 30\% of the nominal value of the installed power has been added to the model. For our case the ramp of the DERs in aggregate form is a total of 0.45 p.u. As can be seen, the new inter-temporal constraint significantly reduces the flexible area in each period. We can notice that the ramping constraint reduces the active power range. In contrast, the reactive power range is not modified because it does not have an associated inter-temporal limitation. 

Finally, in Case III we have added the engineering constraints of voltage and thermal capacities. We can see that the network constraints modify the shape of the flexible regions. In this case, the upper part of the regions is constrained by the capacity of the transmission lines. This means that DERs cannot export all available reactive/active power due to the maximum currents per transmission line constraints.

\subsection{Objective function comparison}

\begin{table}[htbp]
  \centering
  \caption{Surveyor vs. Linear objective functions. Comparison of areas and performance.}
    \begin{tabular}{lccc}
\cmidrule{2-3}    \textbf{CASE I } & Surveyor OF & Linear OF &  \\
\cmidrule{1-3}    Area ($t=1$) & 2.924 (100\%) & 2.924 (100\%) & No \\
    Area ($t=2$) & 3.078 (100\%) & 3.074 (99.9\%) & Ramp \\
    Performace ($s$) & 26.32 & 8.98 &  \\
\cmidrule{1-3}        &     &     &  \\
\cmidrule{2-3}    \textbf{CASE II} & Surveyor OF & Linear OF &  \\
\cmidrule{1-3}    Area ($t=1$) & 2.138 (73.1\%) & 2.143 (73.3\%) & Ramp: \\
    Area ($t=2$) & 2.292 (74.5\%) & 2.289 (74.4\%) & 50\% \\
    Performace ($s$) & 35.6 & 15.27 &  \\
\cmidrule{1-3}    Area ($t=1$) & 1.831 (62.5\%) & 1.827 (62.6\%) & Ramp: \\
    Area ($t=2$) & 1.976 (64.2\%) & 1.977 (64.2\%) & 30\% \\
    Performace ($s$) & 29.94 & 13.93 &  \\
\cmidrule{1-3}    Area ($t=1$) & 1.441 (49.3\%) & 1.441 (49.3\%) & Ramp: \\
    Area ($t=2$) & 1.584 (51.5\%) & 1.584 (51.5\%) & 5\% \\
    Performace ($s$) & 24.19 & 11.15 &  \\
\cmidrule{1-3}        &     &     &  \\
\cmidrule{2-3}    \textbf{CASE III} & Surveyor OF & Linear OF &  \\
\cmidrule{1-3}    Area ($t=1$) & 1.215 (41.5\%) & 1.195 (40.9\%) & Ramp: \\
    Area ($t=2$) & 1.249 (40.6\%) & 1.232 (40\%) & 50\% \\
    Performace ($s$) & 225.3 & 12.24 &  \\
\cmidrule{1-3}    Area ($t=1$) & 0.91 (31.1\%) & 0.901 (30.8\%) & Ramp: \\
    Area ($t=2$) & 0.989 (29.6\%) & 0.972 (31.6\%) & 30\% \\
    Performace ($s$) & 217.64 & 9.73 &  \\
\cmidrule{1-3}        &     &     &  \\
\cmidrule{2-3}    \textbf{CASE IV} & Surveyor OF & Linear OF &  \\
\cmidrule{1-3}    Area ($t=1$) & 1.341 (45.8\%) & 1.464 (50.1\%) & Ramp: \\
    Area ($t=2$) & 1.999 (64.9\%) & 1.5 (48.7\%) & 50\% \\
    Performace ($s$) & 1350 & 118.3 &  \\
\cmidrule{1-3}    Area ($t=1$) & 1.005 (34.4\%) & 1.103 (37.7\%) & Ramp: \\
    Area ($t=2$) & 1.582 (32.6\%) & 1.184 (38.5\%) & 30\% \\
    Performace ($s$) & 1270 & 101.8 &  \\
\cmidrule{1-3}    \end{tabular}%
  \label{tab:of_comparison}%
\end{table}%

The proposed multiperiod model \eqref{eq. model_FR} uses a non-convex formulation of the power flow of a distribution network. We can observe the non-convexity in constraint \eqref{eq:app_power} where there is a bilinear product of the current and voltage variables. NLP solvers (i.e. Ipopt) can solve these bilinear products or Gurobi \cite{optimization2014inc}, which in its latest version can deal with this kind of non-convexities. Additionally, we have modeled two objective functions to maximize the area of the flexibility region. One function seeks to maximize the slopes of specific defined search directions—another objective function based on Surveyor formula to maximize the area. 

\Cref{tab:of_comparison} shows a comparison of the accuracy and performance of the two implemented objective functions. In \Cref{tab:of_comparison}, we evaluate the four cases described in Section B. In the \Cref{tab:of_comparison}, we report the area computed for each period using each objective function. We take as a reference the area calculated with the Surveyor formula without restrictions to determine the relative area in each case. As can be seen in Case I, the Surveyor formulation and the linear formulation have the same area of multiperiod flexibility. However, the non-convexities within the Surveyor formulation make the performance of this model slower. 

For Case II, the areas calculated with the two formulations are comparable. We can conclude that the linear formulation is competitive when we consider ramp constraints. Furthermore, we can see that as the ramp becomes more restricted, the overall flexibility of the system is significantly reduced. Finally, in Case III, the entire model is analyzed with network constraints. Thus, we can conclude that the linear formulation misses some accuracy concerning the Surveyor formula. However, the performance of the linear function is significantly faster than the Surveyor formulation. For Case III, only the ramp levels of 50\% and 30\% were analyzed. In Case IV, it can be analyzed that the entry of batteries increases the flexibility region of the system compared to the area obtained in Case III. This increase in flexibility occurs for both constraint scenarios evaluated. However, storage device modeling significantly increases the complexity of the problem. We can notice that the solution times increase since the new model includes binary variables, making the model MINLP. 

\subsection{Multi-period flexibility region}

\begin{figure}[htbp]
    \centering
    \includegraphics[width=0.8\columnwidth]{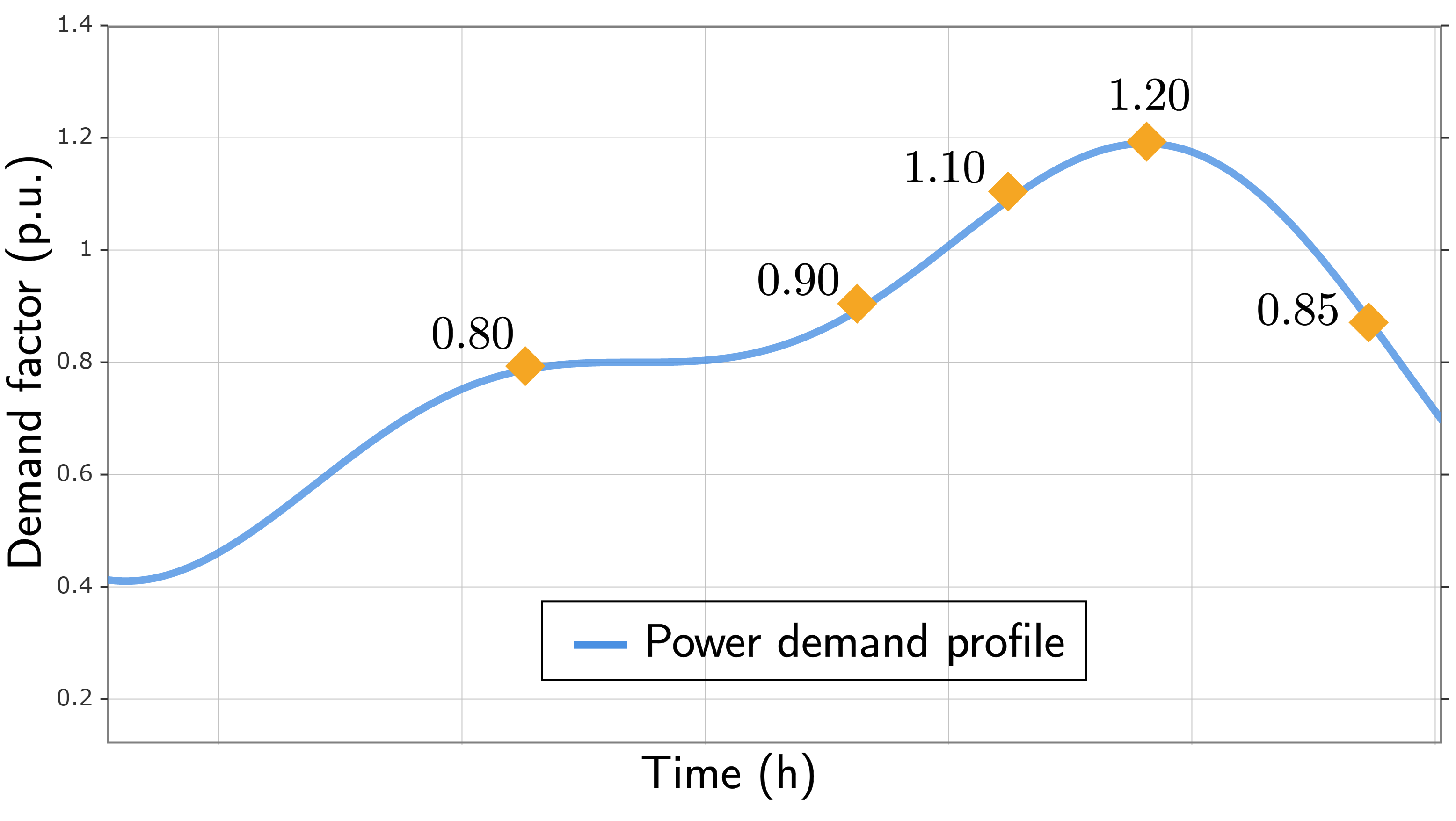}
    \caption{Multiplication factors of the synthetic demand profile. }
    \label{fig:demand}
\end{figure}

The novelty of the proposed methodology lies in the ability to estimate the entire perimeter of the flexibility area through a non-iterative and multi-period approach. The model includes inter-temporal constraints and demand profiles over a given time horizon. Previously, we explained the modifications that the flexibility region experiences in an illustrative way due to changes in the model constraints. In this section, we fully evaluate the 5-node system. For this purpose, we create a synthetic demand profile. The demand profile is shown in \Cref{fig:demand} and the marked values are factors by which the nominal demand is multiplied. This demand profile simulates an aggregate residential demand and a forecast horizon of 5 periods. 

\begin{figure}[htbp]
    \centering
    \includegraphics[width=0.65\columnwidth]{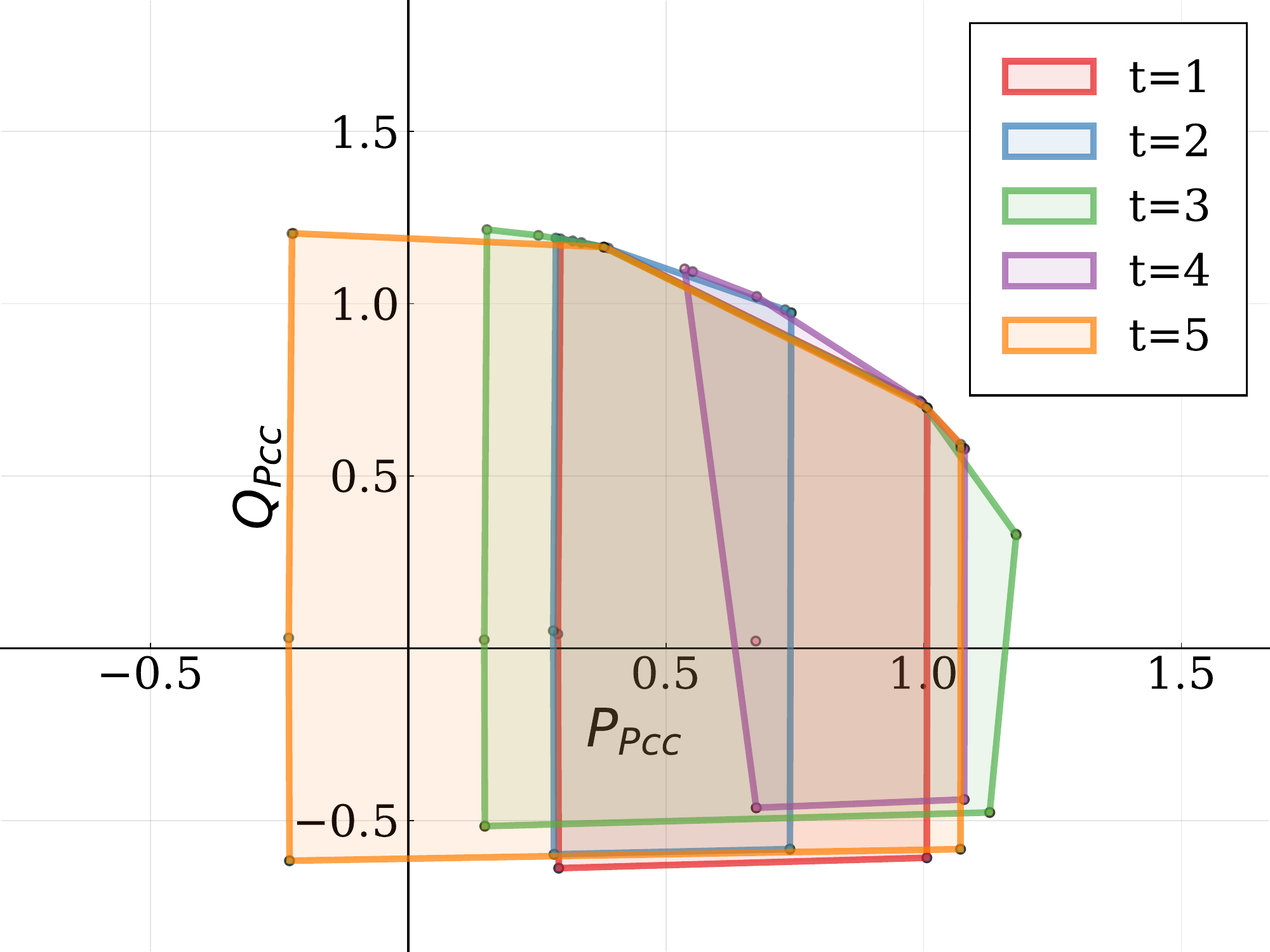}
    \caption{Flexibility map with batteries and distributed generators included.}
    \label{fig:map_region}
\end{figure}

The result obtained from the model with batteries and distributed generators is shown in \Cref{fig:map_region}. The solution flexibility map for the demand periods represents the possible operating points of the feeder substation. Each point in the PQ plane meets the engineering constraints, the inter-temporal constraints and the power balance. As can be seen when energy storage elements are introduced, the multi-period regions increase their flexibility, and their inter-temporal relationship becomes more complicated. The focus of our approach lies in the construction of flexibility maps which are the set of flexible areas across the analysis horizon. 

\begin{figure}
    \centering
    \includegraphics[width=0.8\columnwidth]{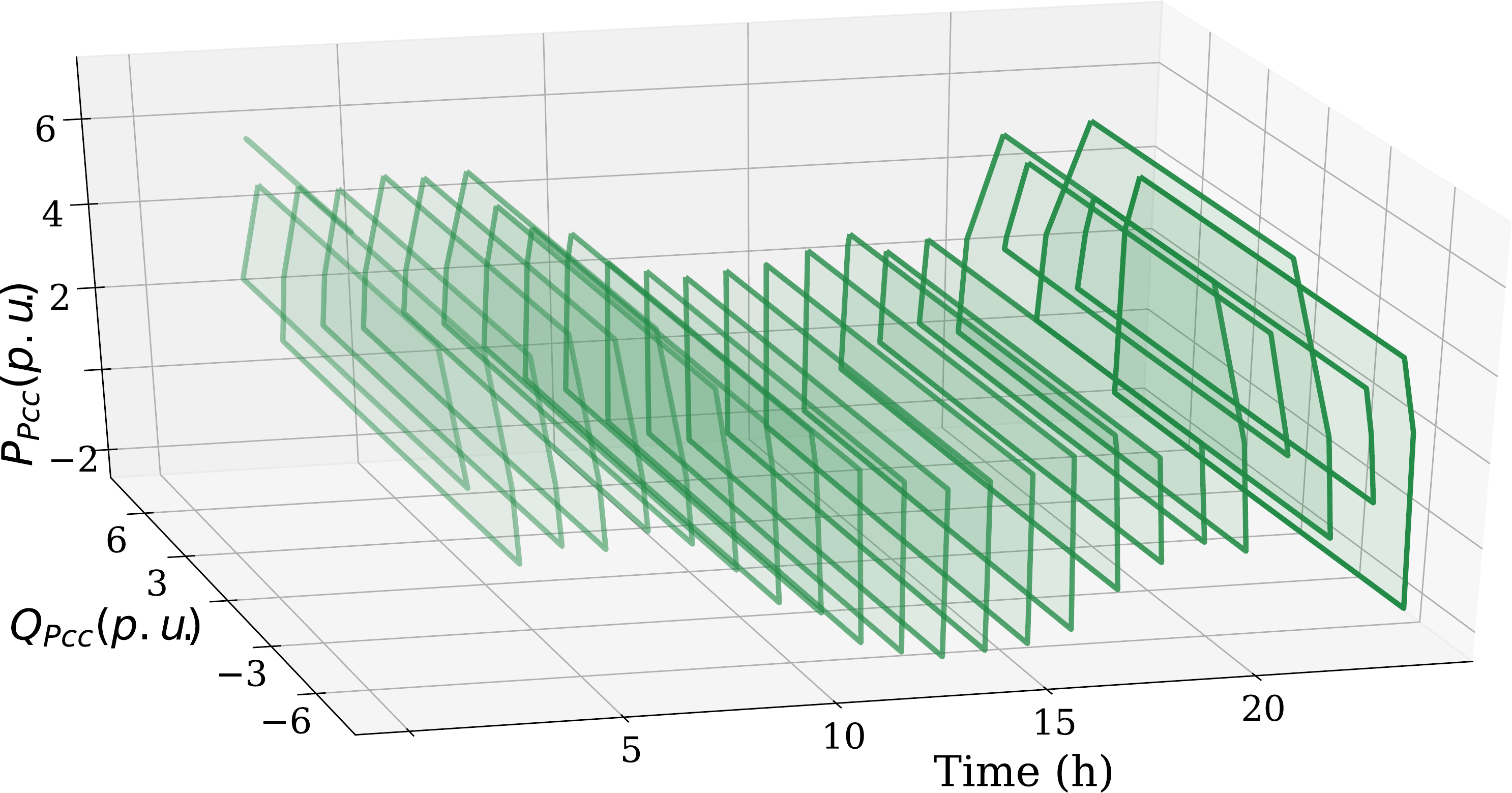}
    \caption{Multi-period flexible region for IEEE 141-node distribution system.}
    \label{fig:Reg-flex-141}
\end{figure}

Finally, we evaluate the scalability of the proposed model on the IEEE 141-node reference system. The distribution network topology and power demand data are the same as in \cite{Khodr2008}. The total system demand is 14,052 MW and 7,410 MVAR which corresponds to a power facor of 0.85. We have added a total of 10 randomly located flexibility elements to the system. We evaluated a total of 24 periods, and the result of the multiperiod flexible region is shown in \Cref{fig:Reg-flex-141}. The figure shows that every period has a flexible region that varies according to network constraints, DER constraints, and hourly demand conditions. 

\section{Discussion} \label{Sec:Discussion}

DERs in an active distribution network increase flexibility and result in a larger area of operation for TSO/DSO coordination. The proposed model results in a set of flexibility regions corresponding to different periods that we can call a flexibility map. The shape of the region changes depending on the PQ capabilities of the flexible resources. The multiperiod flexible region also depends on the ramp rates and the state of the participating DER at time instant $t$. If the DER constraints are not intertemporal, flexibility regions can accurately compute independently. However, the presence of multi-period binding constraints impacts the total potential flexibility available at the TSO/DSO interface. 

The information contained in the flexibility map can be an input to the TSO and used at the daily economic dispatch. The flexibility map estimates the scheduled operating points that the TSO may have. In addition, it shows the possible movements of operating points without violating any technical constraints. The flexibility regions present a set of feasible operations that can aid decision-making in day-to-day operations. It is important to note that the flexibility maps do not provide the combinations of flexibility (the set points of DERs) needed to achieve a specific operating variation. Instead, they show whether that variation can be achieved within the technical limits of the network and DERs. 


\section{Conclusiones} \label{Sec:Conclusions}

The massive entry of DERs creates opportunities for the DSO to provide ancillary services to the TSO. In this paper, the multi-period flexibility region of DERs in an active distribution network is estimated. The model is based on the optimal power flow of distribution networks, and two objective functions are evaluated. The Surveyor formula calculates the maximum area of the multi-period flexible region but has a high computational cost. On the other hand, the maximal P-Q for given directions calculates a decent approximation of the maximum area with a manageable computational cost. An analysis of the impact of inter-temporal constraints on the flexibility region is performed, and it is concluded that they reduce the total flexible region. In comparison, the network constraints modify the region's shape and can limit the flexibilities available in the DERs. The proposed model for calculating the multi-period flexible region is wholly evaluated in a five-node network with DERs (batteries + distributed generators). Finally, we conclude that the presence of DERs can increase the flexibility of the power system by leveraging the entry of renewable generation plants. 

\bibliographystyle{IEEEtran}
\bibliography{references}

\begin{thebibliography}{10}
\providecommand{\url}[1]{#1}
\csname url@samestyle\endcsname
\providecommand{\newblock}{\relax}
\providecommand{\bibinfo}[2]{#2}
\providecommand{\BIBentrySTDinterwordspacing}{\spaceskip=0pt\relax}
\providecommand{\BIBentryALTinterwordstretchfactor}{4}
\providecommand{\BIBentryALTinterwordspacing}{\spaceskip=\fontdimen2\font plus
\BIBentryALTinterwordstretchfactor\fontdimen3\font minus
  \fontdimen4\font\relax}
\providecommand{\BIBforeignlanguage}[2]{{%
\expandafter\ifx\csname l@#1\endcsname\relax
\typeout{** WARNING: IEEEtran.bst: No hyphenation pattern has been}%
\typeout{** loaded for the language `#1'. Using the pattern for}%
\typeout{** the default language instead.}%
\else
\language=\csname l@#1\endcsname
\fi
#2}}
\providecommand{\BIBdecl}{\relax}
\BIBdecl

\bibitem{rogge2017conceptual}
K.~S. Rogge, F.~Kern, and M.~Howlett, ``Conceptual and empirical advances in
  analysing policy mixes for energy transitions,'' \emph{Energy Research \&
  Social Science}, vol.~33, pp. 1--10, 2017.

\bibitem{Bolfek2021}
\BIBentryALTinterwordspacing
M.~Bolfek and T.~Capuder, ``{An analysis of optimal power flow based
  formulations regarding DSO-TSO flexibility provision},'' \emph{International
  Journal of Electrical Power and Energy Systems}, vol. 131, no. January, p.
  106935, 2021. [Online]. Available:
  \url{https://doi.org/10.1016/j.ijepes.2021.106935}
\BIBentrySTDinterwordspacing

\bibitem{Silva2018}
J.~Silva, J.~Sumaili, R.~J. Bessa, L.~Seca, M.~A. Matos, V.~Miranda,
  M.~Caujolle, B.~Goncer, and M.~Sebastian-Viana, ``{Estimating the Active and
  Reactive Power Flexibility Area at the TSO-DSO Interface},'' \emph{IEEE
  Transactions on Power Systems}, vol.~33, no.~5, pp. 4741--4750, 2018.

\bibitem{ageeva2020coordination}
L.~Ageeva, M.~Majidi, and D.~Pozo, ``Coordination between {TSOs} and {DSOs}:
  Flexibility domain identification,'' \emph{arXiv preprint arXiv:2009.02088},
  2020.

\bibitem{Givisiez2020a}
A.~G. Givisiez, K.~Petrou, and L.~F. Ochoa, ``{A Review on TSO-DSO Coordination
  Models and Solution Techniques},'' \emph{Electric Power Systems Research},
  vol. 189, 2020.

\bibitem{burger2017business}
S.~P. Burger and M.~Luke, ``Business models for distributed energy resources: A
  review and empirical analysis,'' \emph{Energy Policy}, vol. 109, pp.
  230--248, 2017.

\bibitem{nosratabadi2017comprehensive}
S.~M. Nosratabadi, R.-A. Hooshmand, and E.~Gholipour, ``A comprehensive review
  on microgrid and virtual power plant concepts employed for distributed energy
  resources scheduling in power systems,'' \emph{Renewable and Sustainable
  Energy Reviews}, vol.~67, pp. 341--363, 2017.

\bibitem{contreras2021}
D.~A. Contreras and K.~Rudion, ``Computing the feasible operating region of
  active distribution networks: Comparison and validation of random sampling
  and optimal power flow based methods,'' \emph{IET Generation, Transmission \&
  Distribution}, 2021.

\bibitem{silva2018challenges}
J.~Silva, J.~Sumaili, R.~J. Bessa, L.~Seca, M.~Matos, and V.~Miranda, ``The
  challenges of estimating the impact of distributed energy resources
  flexibility on the {TSO/DSO} boundary node operating points,''
  \emph{Computers \& Operations Research}, vol.~96, pp. 294--304, 2018.

\bibitem{lopez2021quickflex}
L.~Lopez, A.~Gonzalez-Castellanos, and D.~Pozo, ``Quickflex: a fast algorithm
  for flexible region construction for the tso-dso coordination,'' \emph{arXiv
  preprint arXiv:2107.00114}, 2021.

\bibitem{Riaz2019}
S.~Riaz and P.~Mancarella, ``{On feasibility and flexibility operating regions
  of virtual power plants and TSO/DSO interfaces},'' \emph{2019 IEEE Milan
  PowerTech, PowerTech 2019}, pp. 1--6, 2019.

\bibitem{baran1989optimal}
M.~E. Baran and F.~F. Wu, ``Optimal capacitor placement on radial distribution
  systems,'' \emph{IEEE Transactions on power Delivery}, vol.~4, no.~1, pp.
  725--734, 1989.

\bibitem{braden1986surveyor}
B.~Braden, ``The surveyor's area formula,'' \emph{The College Mathematics
  Journal}, vol.~17, no.~4, pp. 326--337, 1986.

\bibitem{pisciella2017optimal}
P.~Pisciella, M.~T. Vespucci, G.~Vigan{\`o}, M.~Rossi, and D.~Moneta, ``Optimal
  power flow analysis in power dispatch for distribution networks,'' in
  \emph{Numerical Analysis and Optimization}.\hskip 1em plus 0.5em minus
  0.4em\relax Springer, 2017, pp. 229--247.

\bibitem{li2010small}
F.~Li and R.~Bo, ``Small test systems for power system economic studies,'' in
  \emph{IEEE PES general meeting}.\hskip 1em plus 0.5em minus 0.4em\relax IEEE,
  2010, pp. 1--4.

\bibitem{optimization2014inc}
G.~Optimization, ``Inc.,“gurobi optimizer reference manual,” 2015,'' 2014.

\bibitem{Khodr2008}
H.~M. Khodr, F.~G. Olsina, P.~M. O.~D. Jesus, and J.~M. Yusta, ``{Maximum
  savings approach for location and sizing of capacitors in distribution
  systems},'' \emph{Electric Power Systems Research}, vol.~78, no.~7, pp.
  1192--1203, jul 2008.

\end{thebibliography}

\end{document}